\documentclass[11pt]{article}
\usepackage{moriond,epsfig}

\def\Journal#1#2#3#4{{#1} {\bf #2}, #3 (#4)}

\def\NIMA{{\em Nucl. Instrum. Methods} A}

\def\PRL{\em Phys. Rev. Lett.}
\def\PRD{{\em Phys. Rev.} D}

\def\be{\begin{equation}}
\def\ee{\end{equation}}
\def\bea{\begin{eqnarray}}
\def\eea{\end{eqnarray}}

\def\cleo{\mathrm{CLEO}}
\def\cleothree{\mathrm{CLEO~\!III}}
\def\cleoone{\mathrm{CLEO~\!I}}

\def\babar{\mathrm{BaBar}}
\def\belle{\mathrm{Belle}}
\def\bes{\mathrm{BES}}

\def\Y{Y(4260)}
\def\X{X(3872)}
\def\psipp{\psi(3770)}
\def\psip{\psi(2S)}
\def\qqbar{q\overline{q}}
\def\ccbar{c\overline{c}}
\def\deeo{D^{0}}
\def\dstaro{D^{*0}}

\def\threeDone{^{3}D_{1}}
\def\piz{\pi^0}
\def\pip{\pi^+}
\def\pim{\pi^-}
\def\jpsi{J/\psi}
\def\epem{e^{+}e^{-}}

\def\jpc{J^{PC}}
\def\onemm{1^{--}}
\def\onepp{1^{++}}

\def\gev{\!\mathrm{GeV}}
\def\mev{\!\mathrm{MeV}}
\def\kev{\!\mathrm{keV}}
\def\picob{\!\mathrm{pb}}
\def\mevcc{\!\mathrm{MeV}/c^{2}}
\def\bellx{B\rightarrow K X, X\rightarrow\pip\pim\jpsi}

\def\ymass{4284^{+17}_{-16}\pm4~\mevcc}
\def\ywidth{73^{+39}_{-25}\pm5~\mevcc}
\def\oldbind{-0.9\pm2.1~\mev}
\def\newbind{+0.6\pm0.6~\mev}
\def\newdomass{1864.847\pm0.150\pm0.095~\mevcc}
\def\psippjzero{172\pm30}
\def\psippjone{70\pm17}
\def\psippjtwo{21}
\def\chica{75.1\pm3.5\pm4.3\%}
\def\chicb{14.4\pm3.1\pm1.9\%}
\def\chicc{10.5\pm2.4\pm1.2\%}
\def\deut{(3.36\pm0.23\pm0.25)\times10^{-5}}
\def\qqdeut{0.031~\picob}

\begin{document}
\vspace*{4cm}
\title{Bottomonium and Charmonium at $\cleo$}

\author{ R.E. Mitchell \\ (for the CLEO Collaboration)}

\address{Department of Physics, Indiana University, \\
Bloomington, Indiana 47405, USA}

\maketitle\abstracts{
The bottomonium and charmonium systems have long proved to be a rich source of QCD physics.  Recent CLEO contributions in three disparate areas are presented:  (1) the study of quark and gluon hadronization using $\Upsilon$ decays; (2) the interpretation of heavy charmonium states, including non-$c\overline{c}$ candidates; and (3) the exploration of light quark physics using the decays of narrow charmonium states as a well-controlled source of light quark hadrons.}

\section{Introduction}

The $\cleo$ experiment at the Cornell Electron Storage Ring (CESR) is uniquely situated to make simultaneous contributions to both the bottomonium and charmonium systems in a clean $\epem$ environment.  Between 2000 and 2003 $\cleothree$~\cite{cleothree} ran with $\epem$ center of mass energies in the $\Upsilon$ region.  A subset of this period was spent below $B\overline{B}$ threshold, where $\approx$20M, $\approx$10M, and $\approx$5M decays of the $\Upsilon(1S)$, $\Upsilon(2S)$, and $\Upsilon(3S)$, respectively, were collected.  In 2003, CESR lowered its energy to the charmonium region and the $\cleothree$ detector was slightly modified to become CLEO-c~\cite{cleoc}.  Since that time, there has been an energy scan from 3.97 to 4.26~$\gev$ ($\approx 60~\picob^{-1}$), samples collected at 4170~$\gev$ ($\approx 300~\picob^{-1}$, largely for $D_s$ physics) and the $\psipp$ ($\approx 300~\picob^{-1}$, largely for $D$ physics), and a total of nearly 28M $\psip$ decays have been recorded, only 3M of which have been analyzed.

Three (of many) topics recently addressed by the $\cleo$ collaboration will be discussed below.  The reach is wide: from fragmentation in bottomonium decays, to the interpretation of heavy charmonium states, to the use of narrow charmonium states as a source of light quark hadrons.

\section{Bottomonium and Fragmentation}

The bottomonium system provides many opportunities to study the hadronization of quarks and gluons.  The number of gluons involved in the decay of a bottomonium state can be controlled by the charge-conjugation eigenvalue of the initial state:  the $\Upsilon$ states decay through three gluons; the $\chi_{bJ}$ states decay through two.  In addition, the continuum -- where $\epem\rightarrow\qqbar$ proceeds without going through a resonance -- can be used as a source of quarks.  Thus, particle production can be studied and compared in a number of different environments.

\subsection{Quark and Gluon Fragmentation}

\begin{figure}
\centerline{\psfig{figure=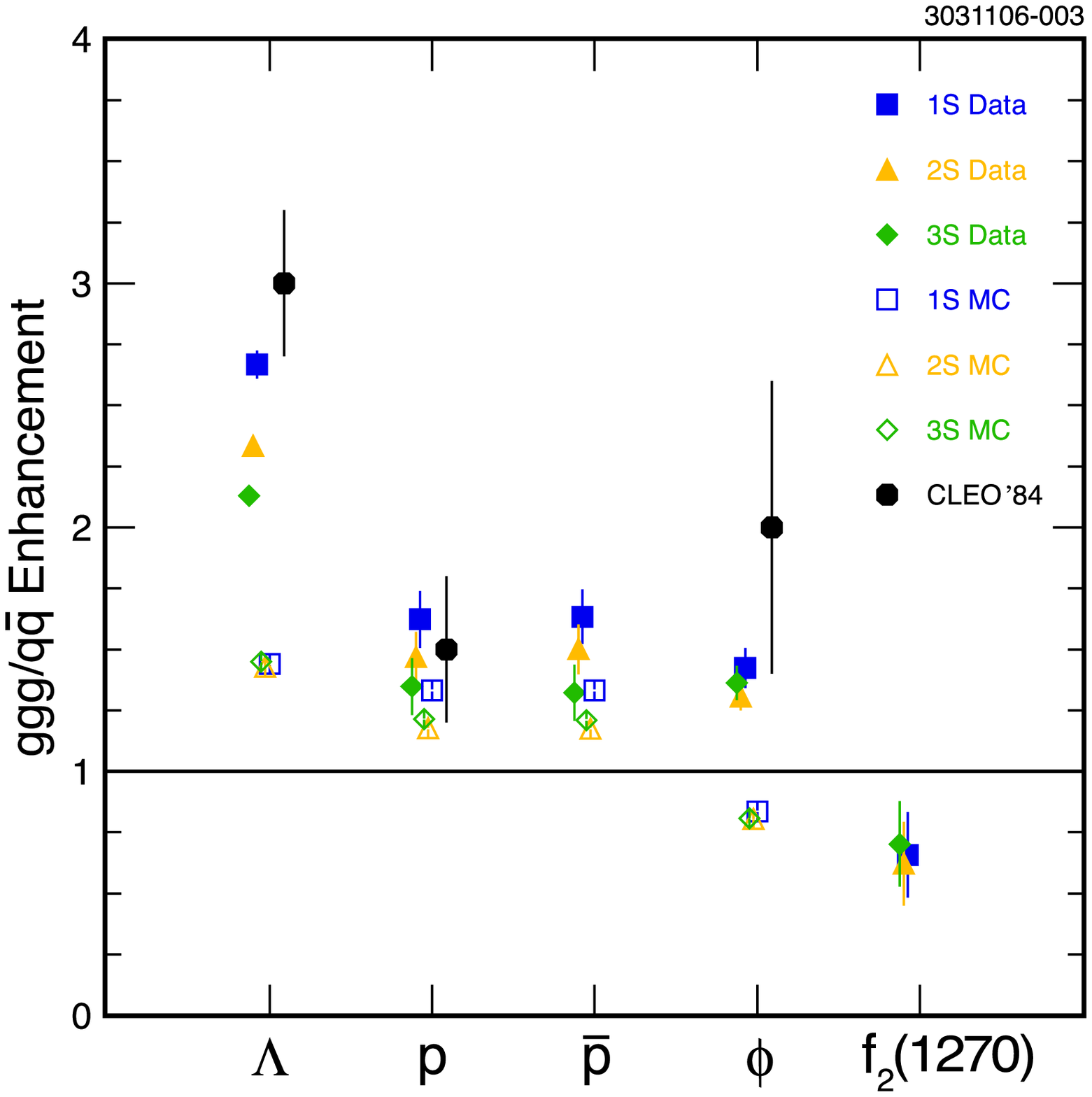,height=2.6in} 
\psfig{figure=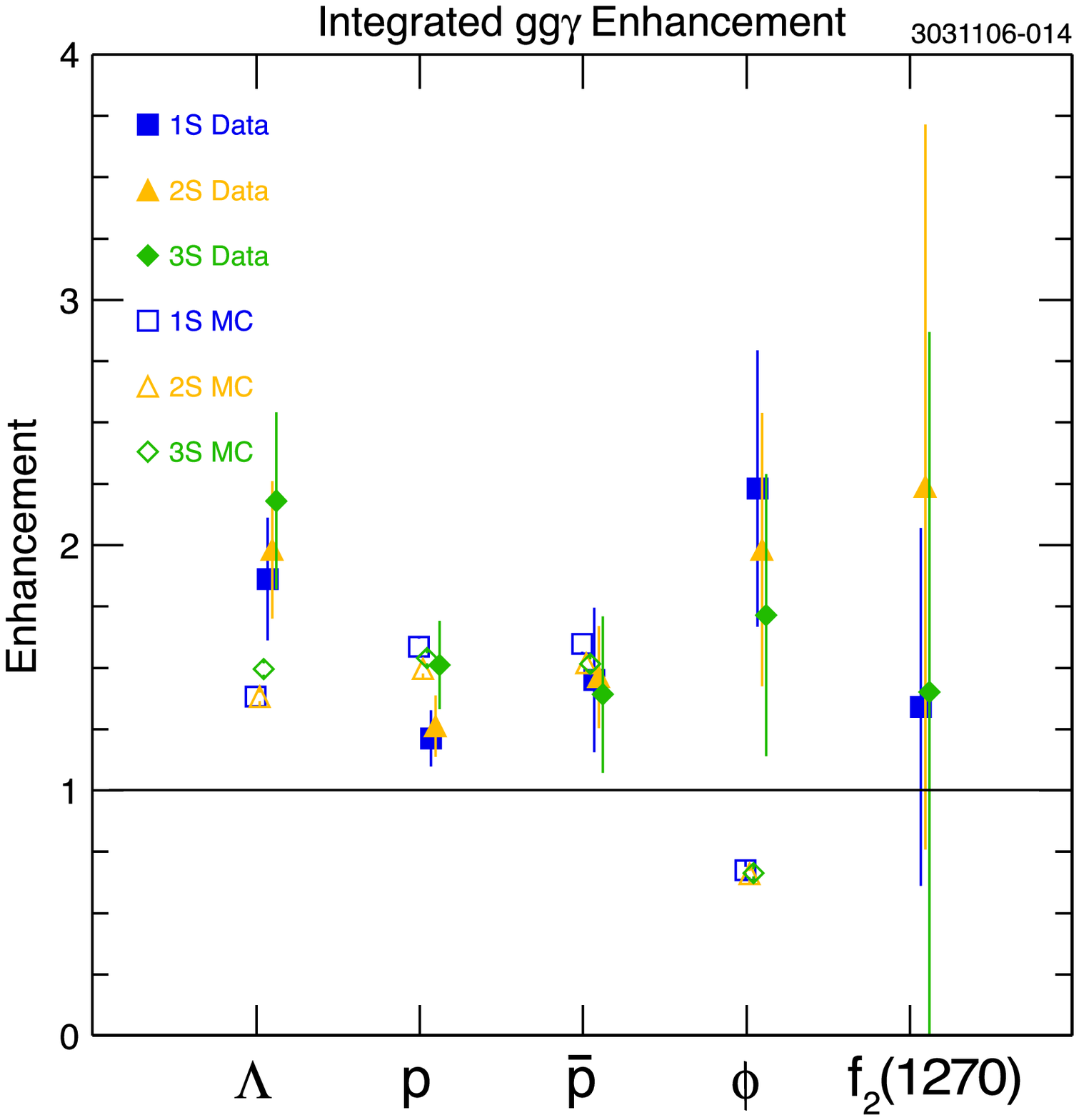,height=2.6in}}
\caption{(a) The enhancement of particle production in $ggg$ ($\Upsilon$ decays) over $\qqbar$ (the continuum).  (b) The enhancement of particle production in $gg\gamma$ (radiative $\Upsilon$ decays) over $\qqbar\gamma$ (radiative continuum events).  See text for details.  From reference~[\ref{ref:ggvsqq}].
\label{fig:frag}}
\end{figure}

In 1984, $\cleoone$ first noticed an enhancement in baryon production in $ggg$ (from $\Upsilon(1S)$ decays) over $\qqbar$ (from the $\epem\rightarrow\qqbar$ continuum), i.e., the number of baryons produced per $\Upsilon(1S)$ decay was greater than the number produced per $\qqbar$ continuum event~\cite{ggold}.  The interpretation of this phenomenon, however, was complicated by the fact that the $ggg$ system consists of three partons (or three strings), while the $\qqbar$ system only has two partons (or one string).  A recent $\cleothree$ analysis~\cite{ggvsqq} has confirmed these findings with greater precision and has extended the comparison beyond $\Upsilon(1S)$ decays to the decays of the $\Upsilon(2S)$ and $\Upsilon(3S)$ states as well.  Figure~\ref{fig:frag}a shows new measurements of the enhancements of particle production in $ggg$ over $\qqbar$, where the ``enhancement'' of a particle species is defined as the ratio of the number of particles produced per event in $\Upsilon$ decays to the number produced per event from the continuum.  The ratio is binned in particle momentum and integrated.   The MC predictions incorporate the JETSET~7.3 fragmentation model.  In addition, the new analysis compares particle production in $gg\gamma$ (radiative $\Upsilon$ decays) and $\qqbar\gamma$ (radiative continuum events).  The comparison in this case is between systems both having two partons and one string.  The energy of the radiated photon is used to monitor parton energies.  Figure~\ref{fig:frag}b shows the enhancements of $gg\gamma$ over $\qqbar\gamma$, where in this case the ratio is binned in the energy of the radiated photon and integrated.  A few conclusions can be drawn from these studies:  (1) baryon enhancements in $gg\gamma$ vs. $\qqbar\gamma$ are somewhat smaller than in $ggg$ vs. $\qqbar$;  (2) the number of partons is important, not just $\sqrt{s}$;  and (3) the JETSET 7.3 fragmentation model does not reproduce the data.

\subsection{Anti-Deuteron Production}


The production of (anti)deuterons in $\Upsilon$ decays provides another opportunity to study the hadronization of quarks and gluons.  In this case, models predict that the gluons from the $\Upsilon$ decay first hadronize into independent (anti)protons and (anti)neutrons, which in turn ``coalesce'' into (anti)deuterons due to their proximity in phase space.  $\cleo$ has measured the production of anti-deuterons in $\Upsilon(1S)$ and $\Upsilon(2S)$ decays and has set limits on their production in $\Upsilon(4S)$ decays~\cite{deut}.  The production of anti-deuterons is easier to measure experimentally than the production of deuterons since anti-deuterons are not produced in hadronic interactions with the detector and the small background makes them easy to spot using $dE/dx$ in the drift chambers.  The relative branching fraction of inclusive $\Upsilon(1S)\rightarrow\overline{d}X$ to $\Upsilon(1S)\rightarrow ggg,gg\gamma$ was found to be $\deut$.  For comparison, a 90\% C.L. upper limit of anti-deuteron production in the continuum was set at $\qqdeut$ at $\sqrt{s}=10.5~\gev$, which, given an hadronic cross-section of the continuum of around 3000~$\picob$, results in less than 1 in $10^{5}$ $\qqbar$ events producing an anti-deuteron.  This is a factor of three less than what is seen in $\Upsilon(1S)$ decays.

\section{Interpretation of Heavy Charmonium States}

The past few years have seen something of a renaissance in charmonium spectroscopy with the discovery of the unexpected $\Y$ and $\X$ states, among others.  The $\Y$ and $\X$, in particular, have been the source of much speculation due to their multiple sightings and the difficulties encountered in attempting to incorporate them into the conventional $\ccbar$ spectrum.  The contributions of $\cleo$ to their interpretation will be discussed below.  In addition, $\cleo$ has recently made measurements pertaining to the charmonium character of the $\psipp$, which is more often used as a source of $D\overline{D}$.  While the $\psipp$ is well-known and has been assumed to be the expected $\threeDone$ state of charmonium, pinning down its properties contributes to our global understanding of the charmonium spectrum.

\subsection{$\Y$}

\begin{figure}
\centerline{\psfig{figure=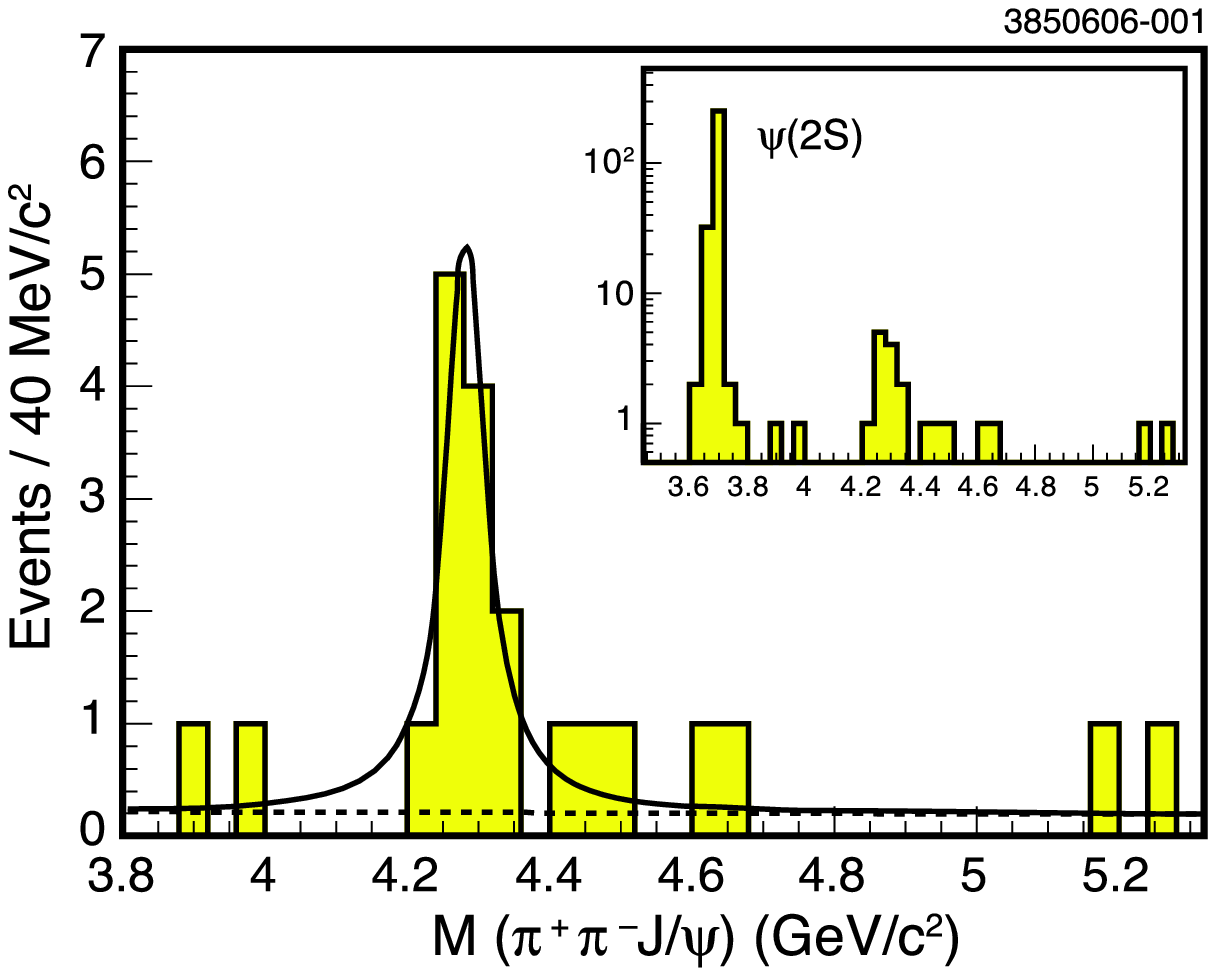,height=2.2in} 
\psfig{figure=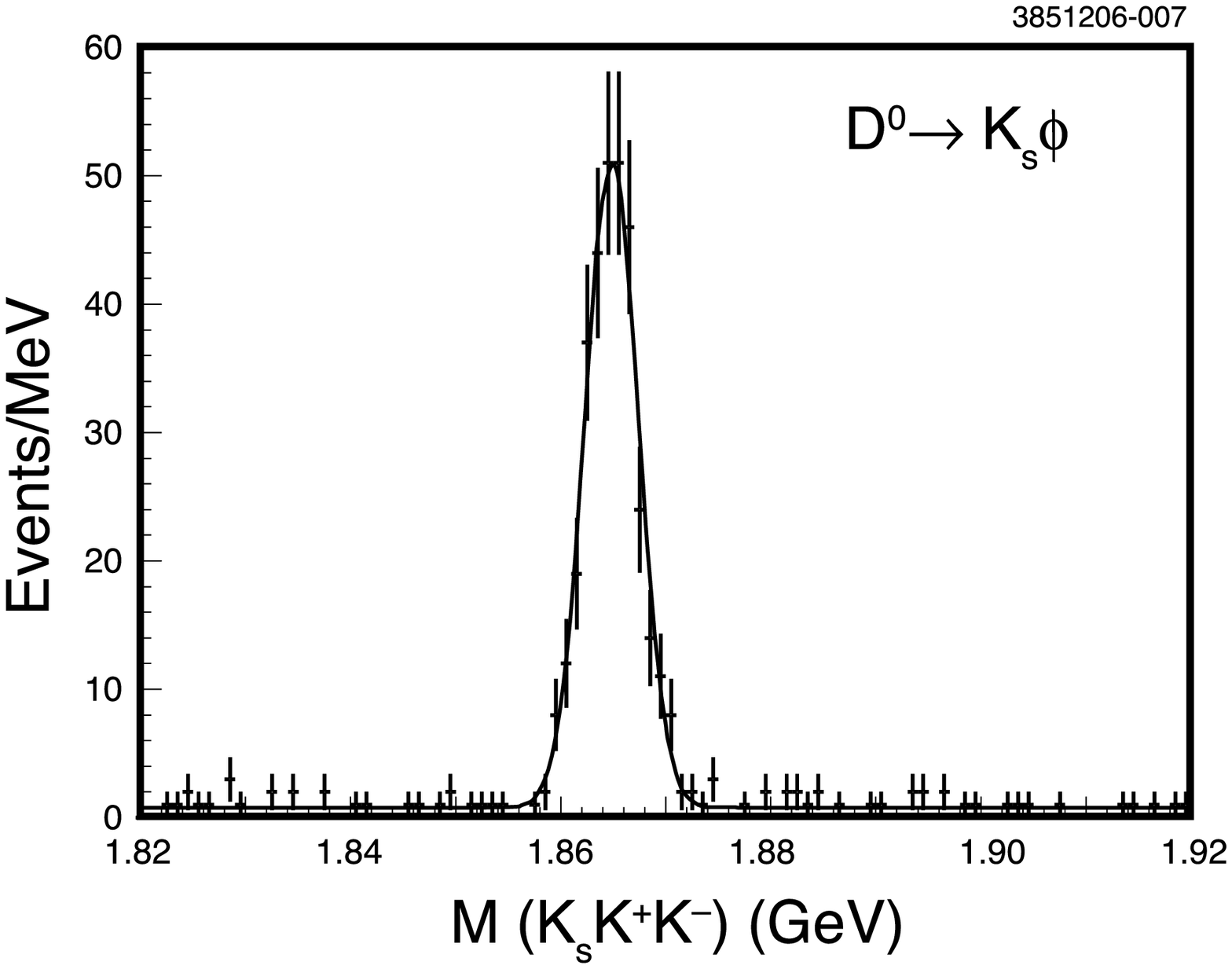,height=2.2in}}
\caption{(a) The ISR production of the $\Y$ in $\cleothree$ (from reference~[\ref{ref:yb}]).  The inset shows the ISR production of the $\psip$.  (b) A precision measurement of the $\deeo$ mass (from reference~[\ref{ref:deeo}]), which aids in the interpretation of the $\X$.
\label{fig:heavy}}
\end{figure}

The $\Y$ was first observed by $\babar$~\cite{ybabar} decaying to $\pip\pim\jpsi$ using $\epem$ collisions with initial state radiation (ISR).  This production mechanism requires the $\Y$ have $\jpc=\onemm$.  However, there is no place for a vector with this mass in the conventional $\ccbar$ spectrum.  On one interpretation the $\Y$ is a hybrid meson, a $\qqbar$ pair exhibiting an explicit gluonic degree of freedom.  $\cleo$ has made two recent contributions regarding the nature of the $\Y$.  First, an $\epem$ energy scan~\cite{ya} was performed between 3.97 and 4.26~$\gev$.  A rise in the production cross section was observed for both $\pip\pim\jpsi$ and $\piz\piz\jpsi$ at 4.26~$\gev$ in the ratio of roughly 2:1.  This ratio suggests the $\Y$ is an isoscalar.  Second, $\cleo$ (using $\cleothree$ data in the $\Upsilon$ region) has confirmed the initial observation by $\babar$ in $\pip\pim\jpsi$ from ISR~\cite{yb} (Figure~\ref{fig:heavy}a).  This both confirms its existence and its $\jpc=\onemm$ nature.  The measured mass and width, $\ymass$ and $\ywidth$, respectively, are also consistent with $\babar$.

\subsection{$\X$}

The $\X$ was first observed by $\belle$~\cite{xbelle} in the reaction $\bellx$.  It has subsequently been studied in several different channels by a variety of different experiments.  From its decay and production patterns it likely has $\jpc=\onepp$.  One of the most tantalizing properties of this state is that its mass is very close to $\deeo\dstaro$ threshold, suggesting that it could be a $\deeo\dstaro$ molecule or a four-quark state.  Prior to the new measurement by $\cleo$, the binding energy of the $\X$ ($M(\deeo)+M(\dstaro)-M(\X)$), assuming it to be a $\deeo\dstaro$ bound state, was $\oldbind$, where the error, perhaps surprisingly, was dominated by the mass of the $\deeo$.  $\cleo$ improved this situation with a new precision $\deeo$ mass measurement~\cite{deeo} using the well-constrained decay $\deeo\rightarrow\phi K_S$ (Figure~\ref{fig:heavy}b) and found the mass to be $\newdomass$.  This results in a small positive binding energy $1~\!\sigma$ from zero: $\newbind$.  This lends further credence to the molecular interpretation of the $\X$.

\subsection{$\psipp$}

\begin{figure}
\centerline{\psfig{figure=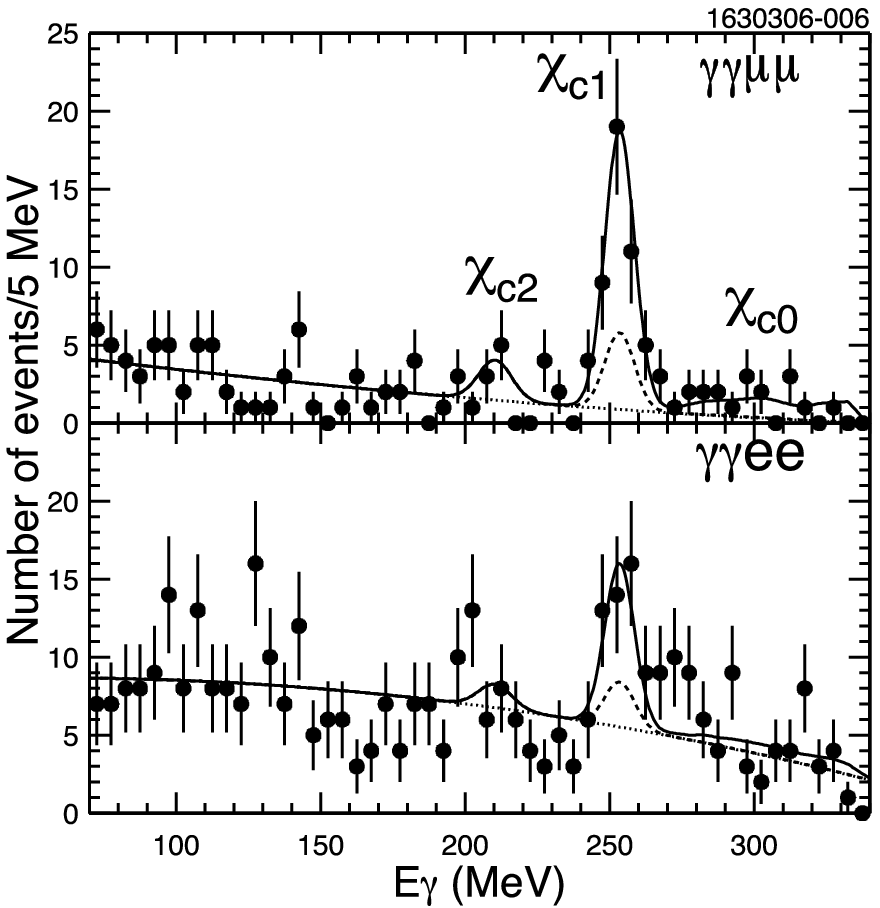,height=2.8in} 
\psfig{figure=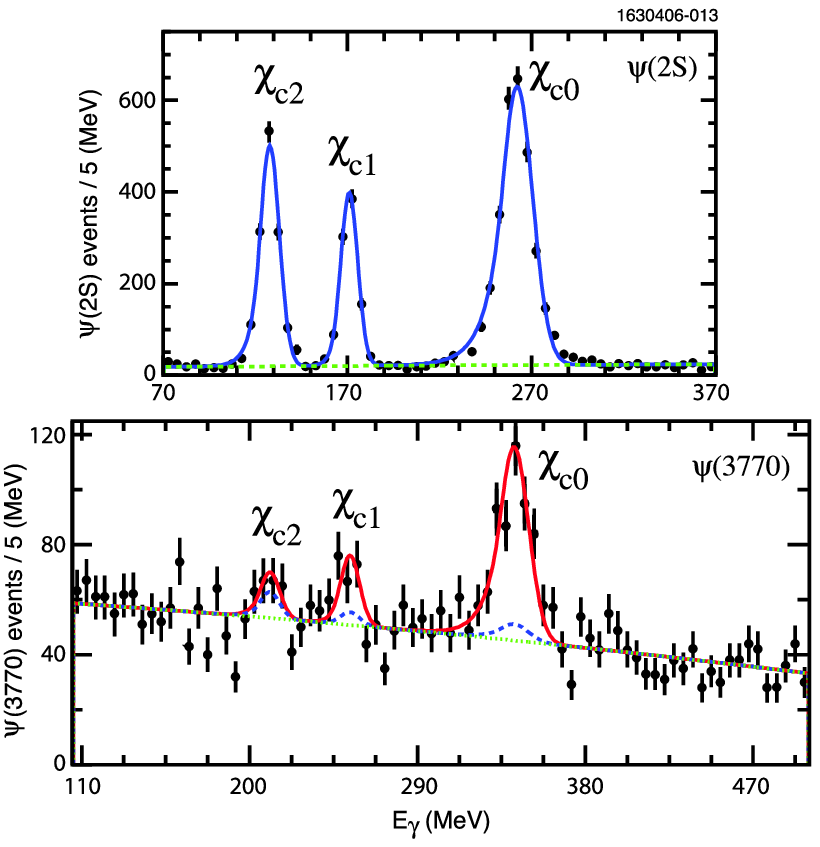,height=2.8in}}
\caption{(a) The energy of the transition photon from $\psipp\rightarrow\gamma\chi_{cJ}$ found when reconstructing $\chi_{cJ}\rightarrow\gamma\jpsi$ and requiring the $\jpsi$ decay to $\mu^+\mu^-$ (top) or $\epem$ (bottom) (from reference~[\ref{ref:psippa}]).  (b) The energy of the transition photon when the $\chi_{cJ}$ are reconstructed in exclusive hadronic modes (bottom).  The top plot shows the same transitions from the $\psip$, which were used for normalization (from reference~[\ref{ref:psippb}]).  The dashed lines in both (a) and (b) are backgrounds from the tail of the $\psip$.
\label{fig:psipp}}
\end{figure}

The existence of the $\psipp$ has been established for a long time.  However, because it predominantly decays to $D\overline{D}$ its behavior as a state of charmonium has been relatively unexplored in comparison to its lighter partners.  The electromagnetic transitions, $\psipp\rightarrow\gamma\chi_{cJ}$, because they are straightforward to calculate,  provide a natural place to study the charmonium nature of the $\psipp$.  $\cleo$ has recently measured these transitions in two independent analyses.  In the first~\cite{psippa}, the processes were measured by reconstructing the $\chi_{cJ}$ in their transitions to $\gamma\jpsi$ and then requiring the $\jpsi$ to decay to $\epem$ or $\mu^{+}\mu^{-}$ (Figure~\ref{fig:psipp}a).  In the second~\cite{psippb}, the $\chi_{cJ}$ were reconstructed in several exclusive hadronic modes and then normalized to the process $\psip\rightarrow\gamma\chi_{cJ}$ using the same exclusive modes (Figure~\ref{fig:psipp}b).  The first method favors the measurement of the transitions to $\chi_{c1}$ and $\chi_{c2}$ while the second method is more suited to the transition to $\chi_{c0}$.  Combining the results of the two analyses, the partial widths of $\psipp\rightarrow\gamma\chi_{cJ}$ were found to be $\psippjzero~\kev$ for $J=0$, $\psippjone~\kev$ for $J=1$, and an upper limit of $\psippjtwo~\kev$ at 90\% C.L. was set for $J=2$.  These measurements are consistent with relativistic calculations assuming the $\psipp$ is the $\threeDone$ state of charmonium.

\section{Using Charmonium to Study Light Quarks}

\begin{figure}
\centerline{
\psfig{figure=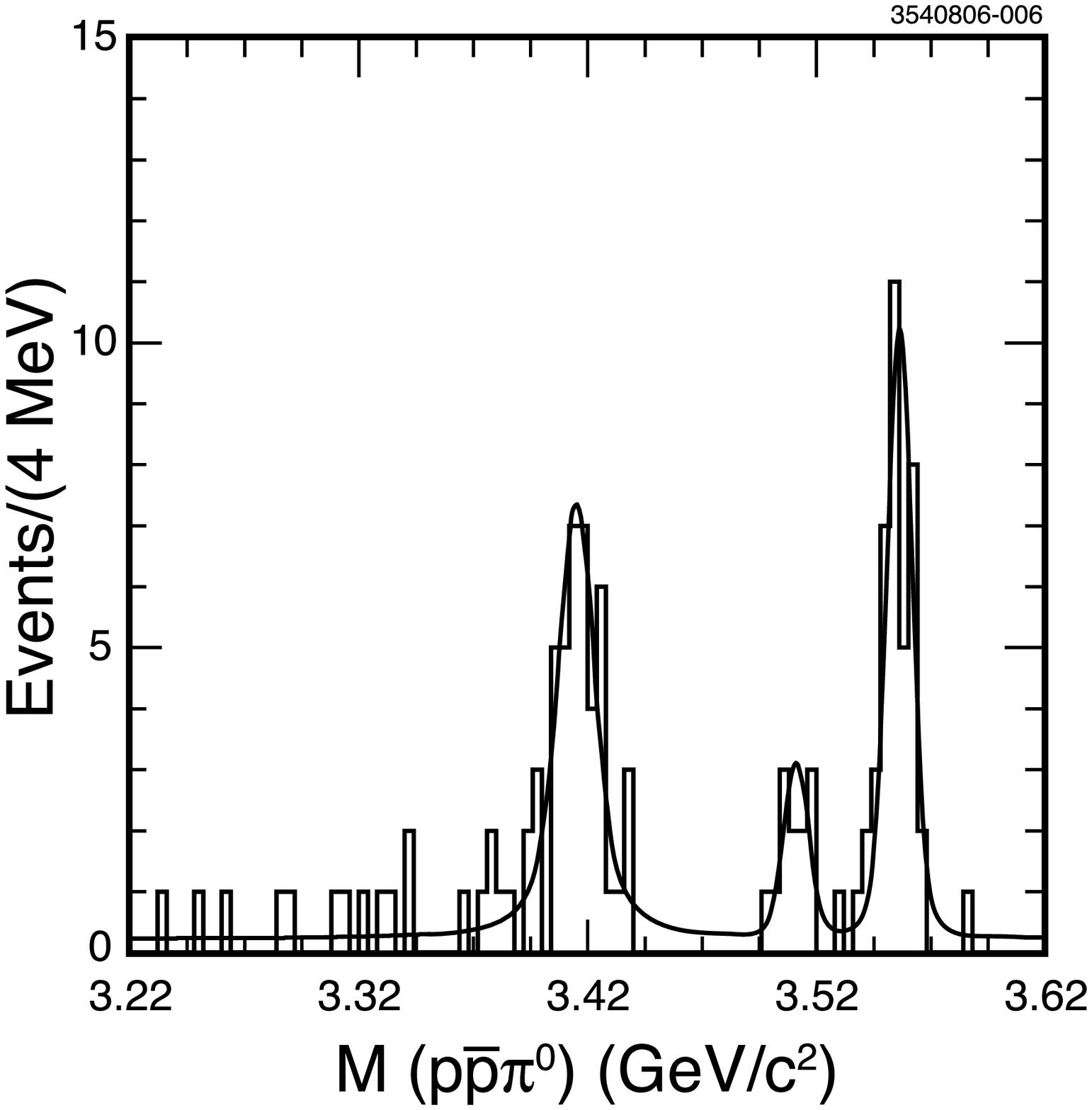,height=2.0in} 
\psfig{figure=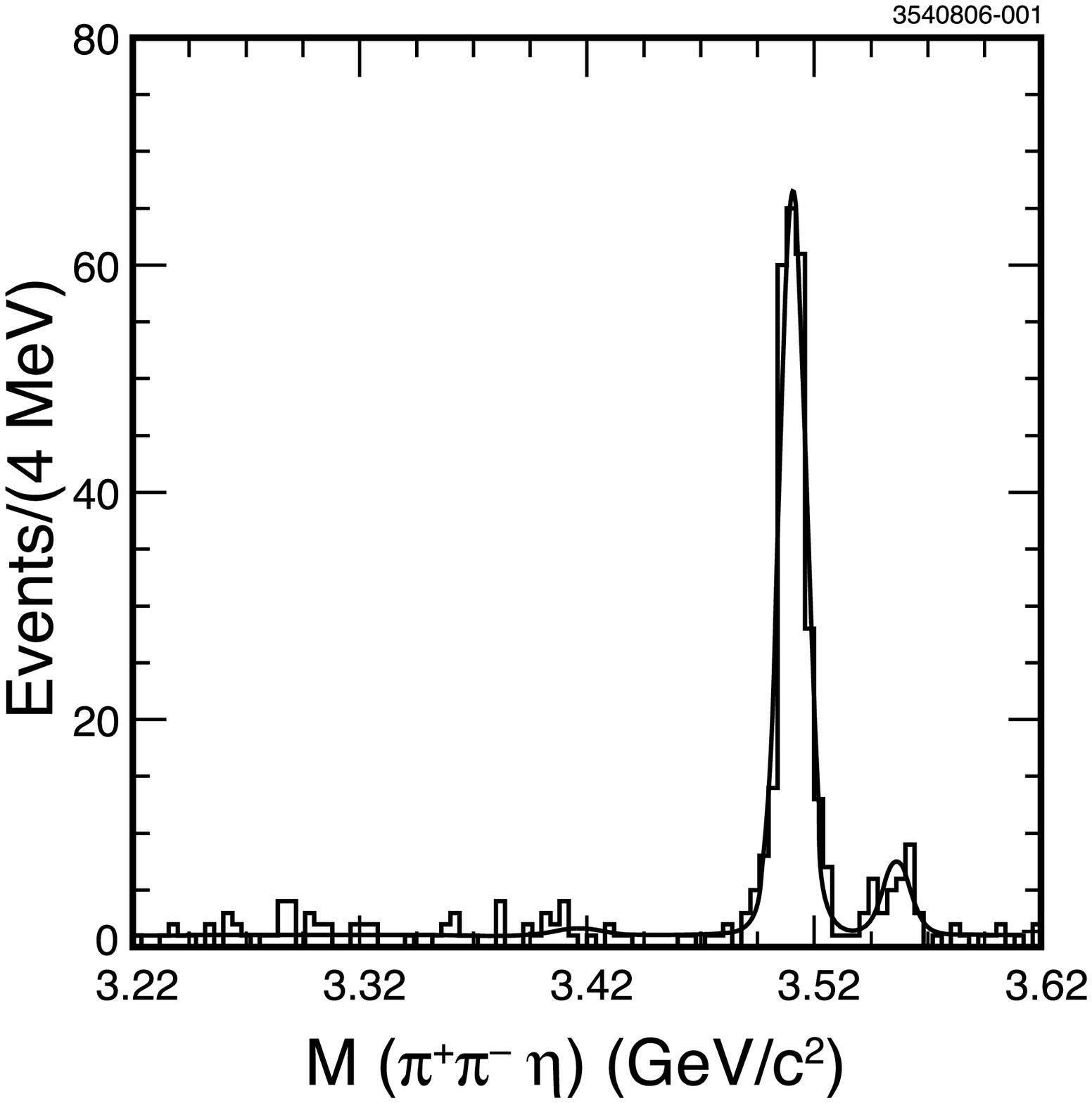,height=2.0in}
\psfig{figure=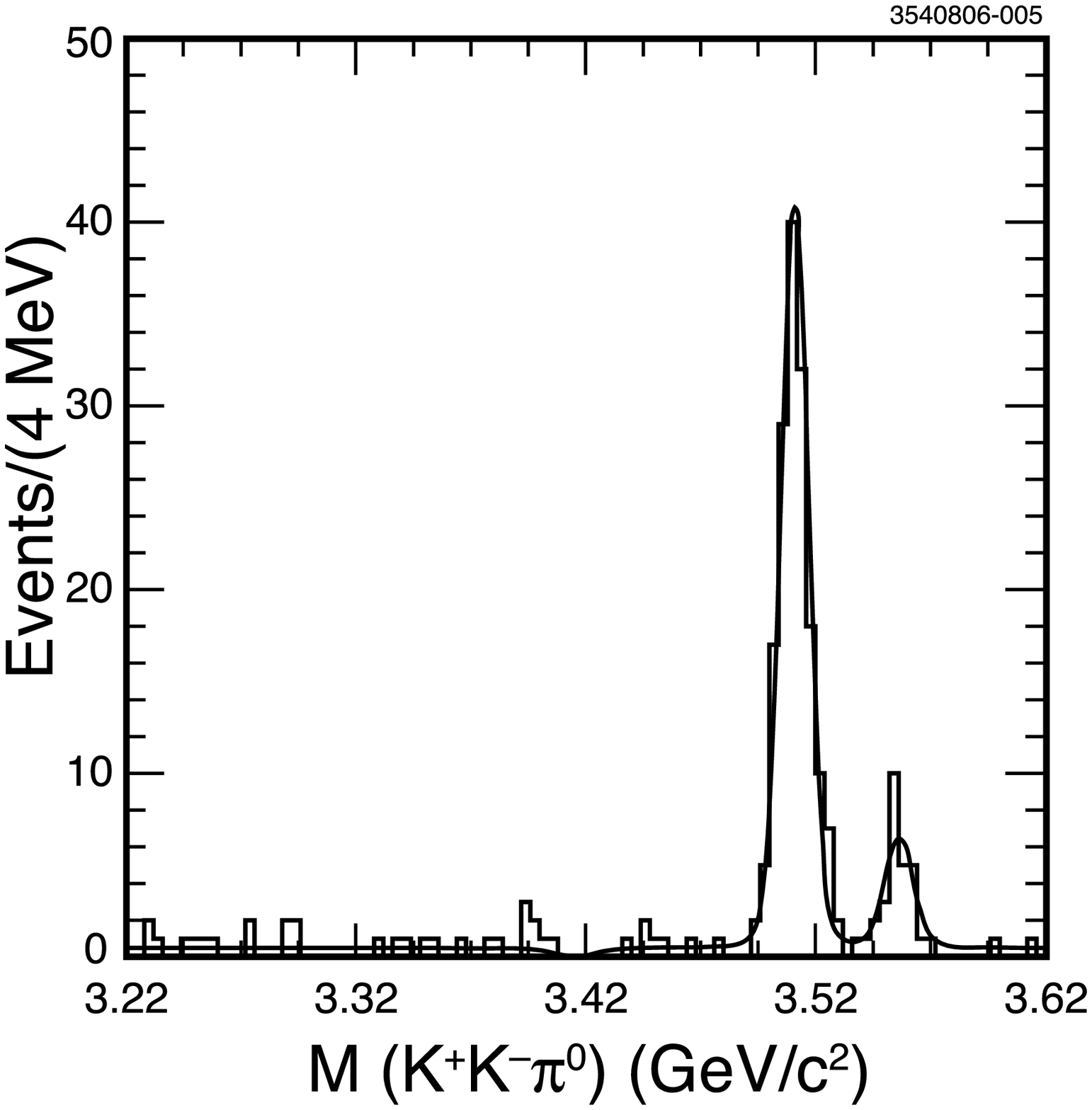,height=2.0in}}
\caption{Reconstructed $\chi_{cJ}$ states ($J=0$, 1, and 2) from the reaction $\psip\rightarrow\gamma\chi_{cJ}$.  From left to right, the $\chi_{cJ}$ states are reconstructed in the exclusive channels $p\overline{p}\piz$, $\eta\pip\pim$, and $K^+K^-\piz$ (from reference~[\ref{ref:chic}]).
\label{fig:chic}}
\end{figure}

In addition to providing valuable information in its own right, the charmonium system can also serve as a well-controlled source of light quark states.  While much effort has gone into the study of $\psip$ and $\jpsi$ decays (e.g. $\jpsi$ radiative decays to glueballs), the decays of the $\chi_{cJ}$ states are less familiar and hold complementary information.  The $\chi_{cJ}$ states are produced proficiently through the reaction $\psip\rightarrow\gamma\chi_{cJ}$, with rates around 9\% for $J=0$, 1, and 2, and can be reconstructed cleanly in many different decay modes in the $\cleo$ detector.

\begin{figure}
\centerline{\psfig{figure=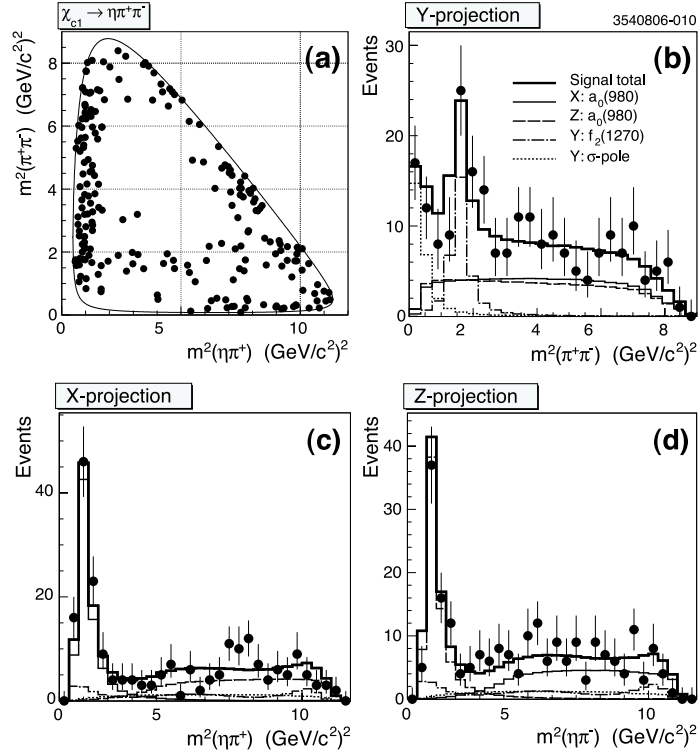,height=4.0in}}
\caption{The Dalitz plot and its projections from the decay $\chi_{c1}\rightarrow\eta\pip\pim$.  Overlaid is a fit to the resonance substructure using a crude non-interfering resonance model (from reference~[\ref{ref:chic}]).
\label{fig:dalitz}}
\end{figure}

As an exploratory study into the analysis of the resonance substructure of $\chi_{cJ}$ decays, $\cleo$ has recently analyzed a series of three-body $\chi_{cJ}$ decays~\cite{chic} using approximately 3M $\psip$ events collected with the $\cleothree$ and CLEO-c detectors.  This anticipates the new sample of approximately 25M $\psip$ events.  The decay modes analyzed include $\eta\pip\pim$, $K^+K^-\eta$, $K^+K^-\piz$, $p\overline{p}\piz$, $p\overline{p}\eta$, $\eta'\pip\pim$, $K_SK^-\pip$, and $K^+\overline{p}\Lambda$.  Branching fractions were measured to each of these final states, many for the first time.  Figure~\ref{fig:chic} shows $\chi_{cJ}$ decays to three particularly well-populated final states.  The $\chi_{c1}$ decays to $\eta\pip\pim$, $K^+K^-\piz$, and $K_SK^-\pip$ included sufficient statistics for a rudimentary Dalitz analysis.  Figure~\ref{fig:dalitz} shows the results of a fit to the $\eta\pip\pim$ Dalitz plot using a crude non-interfering resonance model.  Dominant contributions were found from $a^0(980)\pi$ , $f_2(1270)\eta$, and $\sigma\eta$ with fit fractions of $\chica$, $\chicb$ and $\chicc$, respectively.  No evidence for new structures was found in either $\eta\pip\pim$ or the two $KK\pi$ modes. 

Studies analyzing $\chi_{cJ}$ substructure using the full $\cleo$ sample of 28M $\psip$ decays are underway.  One reaction that looks particularly promising is the decay $\chi_{c0}\rightarrow KK\pi\pi$, which was shown to exhibit a rich substructure of $f$ and $K^*$ states in a recent $\bes$ analysis~\cite{bes}.

\section*{Acknowledgments}
We gratefully acknowledge the effort of the CESR staff in providing us with excellent 
luminosity and running conditions. This work was supported by the National Science 
Foundation and the U.S. Department of Energy.

\end{document}